\documentclass[prl,twocolumn,reprint,showpacs,preprintnumbers,amsmath,amssymb]{revtex4-1}
\usepackage{graphicx}
\usepackage{dcolumn}
\usepackage{bm}
\usepackage{color}
\usepackage{pinlabel}

\newcommand{\no}{\nonumber}

\usepackage{braket}

\begin{document}

\title{Small-q phonon mediated singlet and chiral spin triplet superconductivity in LiFeAs}

\author{A. Aperis}\email{aaperis@mail.ntua.gr}
\author{G. Varelogiannis}\email{varelogi@mail.ntua.gr}
\affiliation{Department of Physics, National Technical University of Athens, GR-15780 Athens, Greece}
\vskip 0.7cm

%71.27.+a    Strongly correlated electron systems; heavy fermions
%74.25.-q 	Properties of superconductors
%74.20.-z    Theories and models of superconducting state
%74.10.+v    Occurrence, potential candidates
%74.20.Rp 	Pairing symmetries (other than s-wave)
%74.70.Xa 	Pnictides and chalcogenides
%74.25.Kc 	Phonons

\begin{abstract}
We report fully momentum dependent, self-consistent calculations of the gap symmetry, Fermi surface (FS) anisotropy and $T_c$ of superconducting (SC) LiFeAs using the experimental band structure and a realistic small-q electron phonon interaction within the framework of Migdal-Eliashberg theory. In the stoichiometric regime, we find the exact $s_{++}$ gap as reported by ARPES. For slight deviations from stoichiometry towards electron doping, we find that a
\emph{chiral triplet} $p_x+ip_y$ state stabilizes near $T_c$ and that at lower temperatures a
\emph{transition from the triplet to singlet} $s_\pm$ SC takes place. Further doping stabilizes the chiral p-wave SC down to T=0. Precisely the same behavior was observed recently by NMR. Our results provide a natural and universal understanding of the conflicting experimental observations in LiFeAs.
\end{abstract}

\maketitle

%\paragraph{Introduction.}
In iron arsenide superconductors are observed
not only the higher critical temperatures after those observed in cuprates but also
a plethora of exciting phenomena yet to be understood.
Perhaps the most challenging compound is the stoichiometric LiFeAs %having a T$_c$=16-18 K without doping or nearby magnetic phases
\cite{lifeasprops} which exhibits a suprisingly exotic superconducting phenomenology.
Measurements of NMR \cite{ZhengLi}, inelastic neutron scattering
 \cite{Boothroyd} and specific heat \cite{DJJang} in this compound have been interpreted in terms of multigap singlet s$_\pm$ SC \cite{Mazin2}.
However, the gap seen by ARPES indicates an anisotropic, singlet and one sign s$_{++}$ order parameter \cite{Bor1,Umezawa}.
On the other hand, there is also consistent evidence for \emph{p-wave spin triplet SC} in LiFeAs, first reported by
quasiparticle interference (QPI) \cite{QPItr} and NMR \cite{Baek}. Subsequent high magnetic field measurements have found that this state can also be induced by the field and it is \emph{chiral} \cite{lifeasField}.

The NMR data have posed yet another extraordinary puzzle.
Depending on small changes of the stoichiometry,
some of the samples exhibit singlet SC, while the slightly more electron doped samples show triplet SC \cite{Baek}. Moreover, it was recently reported that some samples exhibit \emph{a temperature induced singlet to triplet transition} as temperature rises \cite{Baek2}.
Clearly, not only do we observe in LiFeAs the highest temperature, by an order of magnitude, \emph{triplet} SC state
(T$_c$=16-18 K), but in addition this state emerges
away from any magnetically ordered
phase and exhibits some of the most surprising
phase transition phenomena in a SC regime.
Therefore, understanding theoretically the complex and seemingly conflicting
SC phenomenology of
LiFeAs is arguably one of the most challenging issues
in the field of superconductivity.

The electronic structure of LiFeAs is characterized by reduced nesting properties and a \emph{Van Hove} point (VHp) at the \emph{center}
of the Brillouin Zone (BZ) \cite{SCnest} (see Fig.1(a)). Although antiferromagnetic spin fluctuations (SFs) are generally weak in this material
 \cite{ZhengLi,Bor1,Qureshi}, their importance for SC is yet unresolved \cite{Knolle}. On the contrary, there is clear evidence of an enhanced electron-phonon interaction (EPI), strong enough to account for lifetime effects and the order of magnitude of the measured T$_c$ \cite{Kordyuk}.
So far, it has been argued that ferromagnetic SFs may drive the triplet SC \cite{Brydon}. On the other hand, the s$_{++}$ SC has been interpreted in terms of orbital fluctuations assisted by phonons \cite{Kontani}. Even if either mechanism is proven relevant for LiFeAs, it still remains unclear how it could explain coherently the
conflicting gap symmetry observations.

In this Letter we demonstrate that \emph{all} of the aforementioned exotic phenomenology is consistently explained, in a unified way,
if the enhanced EPI in this material has a momentum dependence peaked at \emph{small-q}, similar to the one observed recently in cuprates \cite{smqcupExp}. Our analysis is based on fully anisotropic, self-consistent calculations of the SC gap within the Eliashberg approach, assuming a realistic small-q EPI and using the ARPES resolved multi-band structure for LiFeAs as reported by the experiments.
Remarkably, using an overall electron-phonon coupling (EPC) strength
of $\lambda=1.38$ consistent with phonon damping effects in the ARPES spectra
and reasonable characteristic phonon frequencies
of about 100 K, the same Eliashberg calculations produce
the right critical temperatures as well.

Small-q EPI can occur due to strong ionic and/or Coulombic effects \cite{Pickett,smq1,Weger,Zeyher,Grilli,smqGV1,smqGV2,Huang,smqcupThe}.
Note that large ionic polarizabilities that may lead to enhanced dielectricity have indeed been reported
in pnictides \cite{Sawatzky,Drechsler}.
It was early on understood that small-q EPI may lead to \emph{unconventional}
SC of d-wave type in the cuprates \cite{smqGV2} and recent findings support the relevance of this type of interaction in these
materials \cite{smqcupExp,smqcupThe}. Other unconventional SC materials, including
$\kappa$-BEDT organic salts \cite{SmallQorganic}, 
the heavy fermion UPd$_2$Al$_3$ \cite{SmallQheavy} and cobaltites
\cite{SmallQcobaltites} have also been shown within a BCS framework to be
compatible with this generic picture.
Recently, the first multiband BCS
approach with the small-q
EPI mechanism in pnictides \cite{Aperis} established that the reported
unconventional SC states are compatible with the
measures of finite isotope effect \cite{IsotopeNat}. Our present fully
momentum dependent Eliashberg calculations specifically dedicated to LiFeAs
are of unprecedented precision. The quality with which the experimental
behavior of LiFeAs is reproduced as well as the fact that a \emph{chiral triplet} SC state
in a real material is plausibly associated with
a \emph{phononic} mechanism, has broad implications
for the whole field of unconventional superconductivity.

We parametrize the EPC as an effective, yet realistic, kernel separable over momentum and Matsubara frequency: $\lambda({\bf k},{\bf k}';n,n')$=$\lambda_{{\bf kk}'}\lambda_{nn'}$, where $\lambda_{nn'}$ has an Einstein form and $\lambda_{{\bf k},{\bf k}'}$=$\frac{N_0 V_{ph}}{q^2_c+|{\bf k}-{\bf k}'|^2}$,
with $V_{ph}$ an effective potential and $N_0$ the DOS on the FS. The momentum cutoff $q_c$ selects the small wave vectors in the attractive phonon part while at larger wave vectors the repulsive Coulomb pseudopotential may prevail. For example, such a cutoff may arise naturally due to the interplay between enhanced dielectricity and the long range nature of the Coulomb potential. The simplest approximation to this case maps $q_c$ to the Thomas-Fermi (TF) screening wavevector, $q_{TF}\propto\sqrt{N_0/\epsilon}$ where $\epsilon$ is the dielectric constant \cite{smq1,Weger,smqcupExp}. In the case of LiFeAs, this provides with the rough estimate that $q_c$ should decrease away from stoichiometry, i.e. from the VHp. A small $q_c$ leads to a {\it momentum decoupling} (MD) situation \cite{smqGV1}. In the MD regime the gap function loses its rigidity in momentum space and becomes correlated with the variations of the DOS, leading to FS momentum dependent and possibly unconventional SC. For a multiband system, this also leads to the dominance of the EPI for intraband channels, while the Coulomb pseudopotential provides a repulsive interband coupling, allowing for a sign alternating SC gap \cite{Aperis}.
%This mechanism resembles the one provided by SF for the emergence of s$_{\pm}$ SC \cite{Mazin} and has been shown to lead to similar results \cite{Aperis}.
%The combination of MD phenomena and the repulsive Coulomb pseudopotential provides a unique phononic route to unconventional SC symmetries, as has been often demonstrated in the past \cite{}.
Since by definition: $\lambda=\langle\lambda({\bf k},{\bf k}';0)\rangle_{{\bf k},{\bf k}'}^{FS}$, in the following we fix $\lambda$ to the experimentally reported value \cite{Kordyuk} by adjusting $V_{ph}$, so that for each value of $q_c$, $\lambda=\langle\lambda_{{\bf k},{\bf k}'}\rangle_{{\bf k},{\bf k}'}^{FS}=1.38$, where $\langle\ldots\rangle_{\bf k}^{FS}$ is a FS average.
% over the whole FS.
\begin{figure}[t!]
%\begin{center}
\begin{minipage}[b]{1.105in}
\labellist
\footnotesize\hair 2pt
\footnotesize\pinlabel a) at 20 150
%\scriptsize\pinlabel $\mu$=0.00 at 130 210
\scriptsize\pinlabel $\mu$=0 at 130 15
\tiny\pinlabel $\Gamma$ at 130 111
\tiny\pinlabel X at 244 111
\tiny\pinlabel M at 244 214
\scriptsize\pinlabel $\alpha$ at 150 95
\scriptsize\pinlabel $\beta$ at 185 60
\scriptsize\pinlabel $\gamma$ at 242 20
\scriptsize\pinlabel $\delta$ at 220 40
\endlabellist
\includegraphics[width=\textwidth]{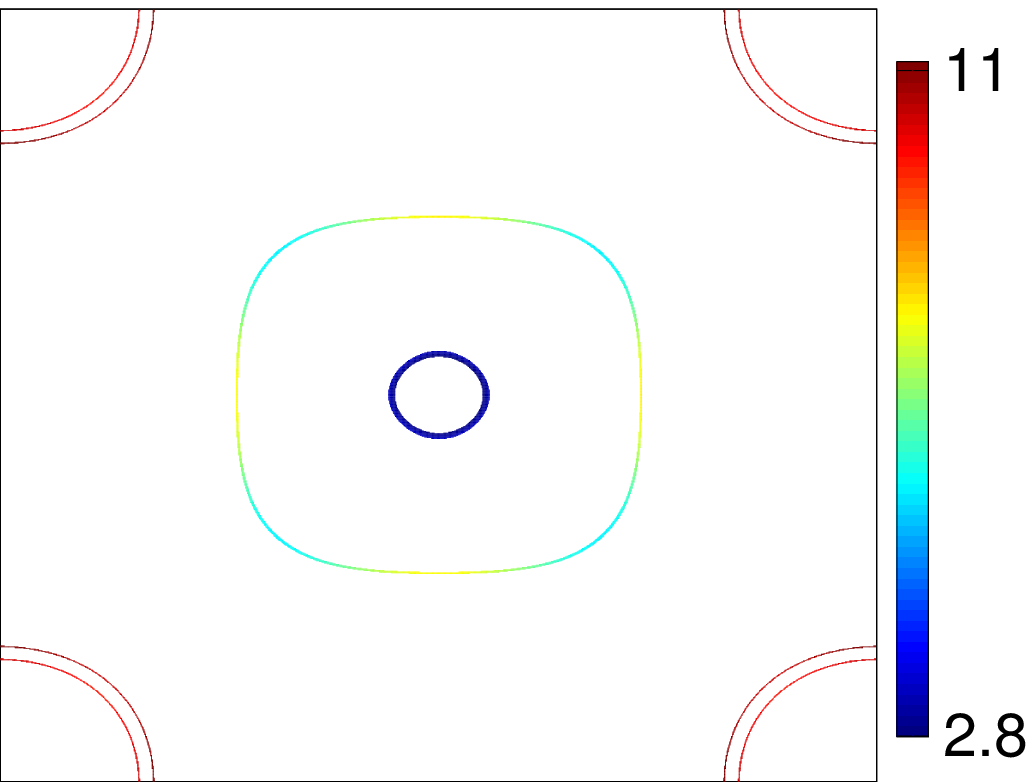}\vspace{0.02in}
\end{minipage}
\begin{minipage}[b]{1.105in}
\labellist
\footnotesize\hair 2pt
\footnotesize\pinlabel b) at 20 150
%\scriptsize\pinlabel $\mu$=0.00 at 130 210
%\scriptsize\pinlabel $\mu$=30meV at 130 15
\scriptsize\pinlabel $\mu$=10meV at 130 15
\tiny\pinlabel $\Gamma$ at 130 111
\tiny\pinlabel X at 244 111
\tiny\pinlabel M at 244 214
\endlabellist
\includegraphics[width=\textwidth]{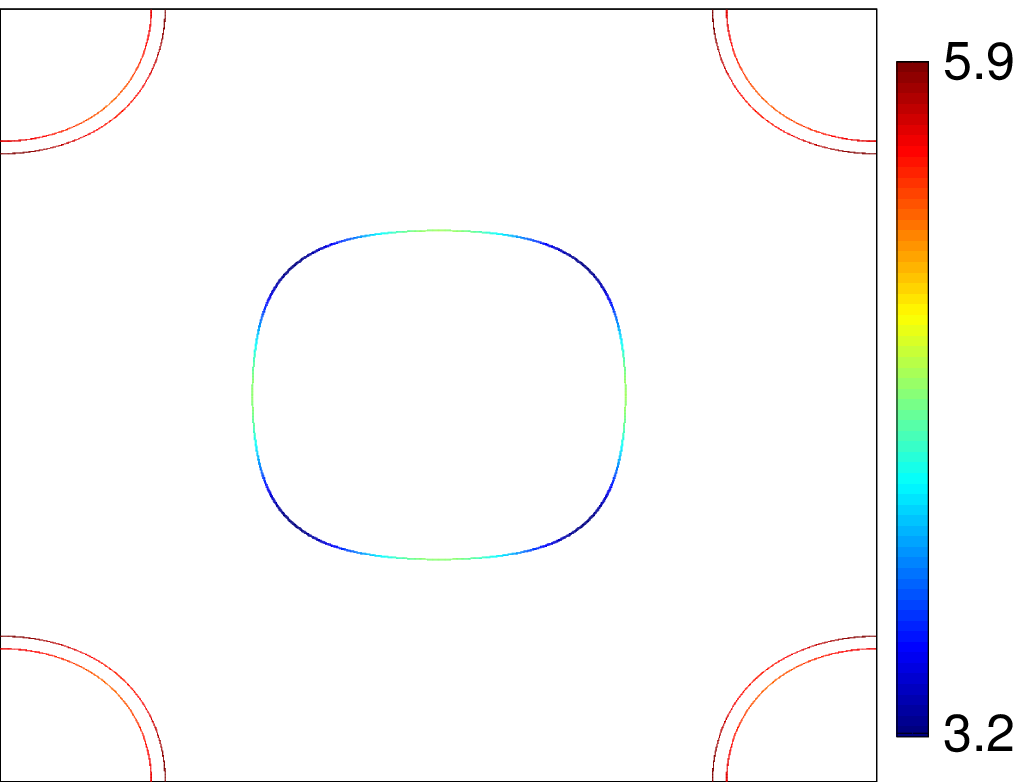}\vspace{0.02in}
\end{minipage}
\begin{minipage}[b]{1.105in}
\labellist
\footnotesize\hair 2pt
\footnotesize\pinlabel c) at 20 150
%\scriptsize\pinlabel $\mu$=0.00 at 130 210
\scriptsize\pinlabel $\mu$=50meV at 130 15
\tiny\pinlabel $\Gamma$ at 130 111
\tiny\pinlabel X at 244 111
\tiny\pinlabel M at 244 214
\endlabellist
\includegraphics[width=\textwidth]{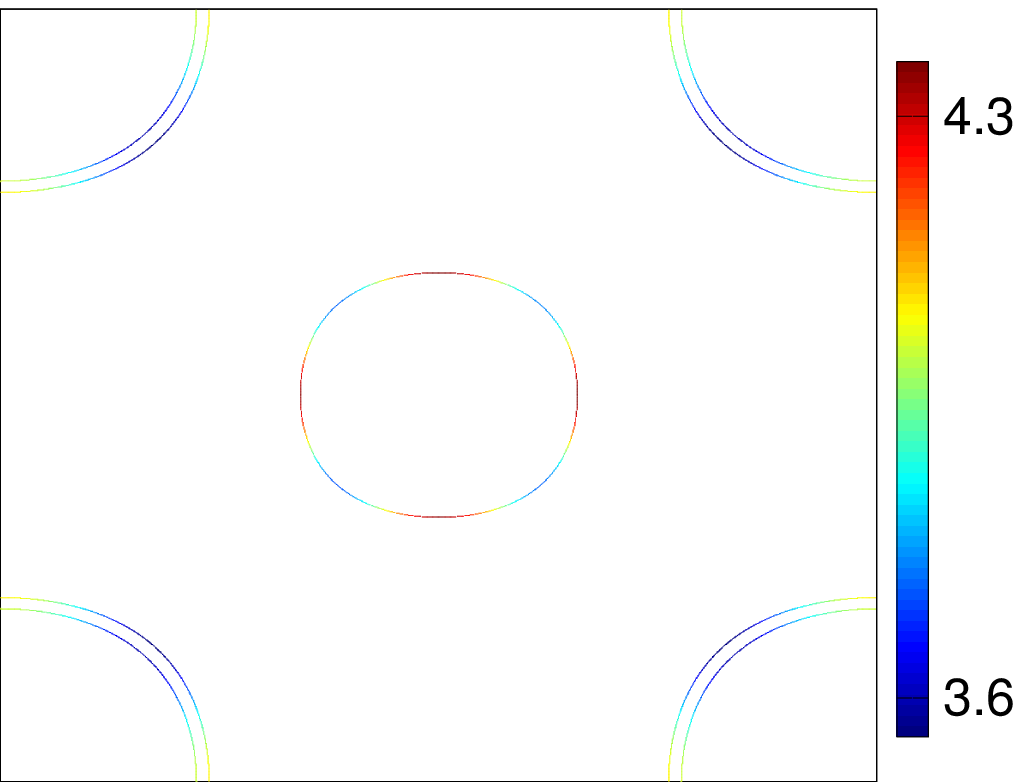}\vspace{0.02in}
\end{minipage}
%\end{center}
\caption{(Color online) (a)-(c) The FS of LiFeAs colored by velocity for $\mu$=$[0,10,50]$meV. The $\alpha$ pocket is very shallow and posseses high and isotropic DOS. The remaining FS pockets, exhibit DOS maxima along $\Gamma$-M that are enhanced by doping.}\label{q3}
\end{figure}

For the band structure of LiFeAs we use a four band Tight-Binding (TB) model elaborated by the IFW group as a fit to ARPES data \cite{Lankau, Knolle}. Hence our realistic input, captures the experimentally resolved FS of LiFeAs extremely well (Fig.\ref{q3}(a)). We model the effect of doping in the rigid band approximation by substracting a chemical potential, $\mu$, from all four bands. For $\mu>0$ ($\mu<0$) the system is electron (hole) doped. We refer to the inner hole, outer hole, inner electron and outer electron pocket as $\alpha$,$\beta$,$\gamma$ and $\delta$, respectively. In Fig.\ref{q3}(a)-(c) the FS of this model is shown, colored by velocity at different values of $\mu$ . The presence of a VHp at $\Gamma$ is evident by the blue color denoting a very high and isotropic DOS over $\alpha$. This pocket is very shallow thus, a slight deviation from stoichiometry, which translates to $\mu>5.2$meV,
%(or 0.01\% electron doping)
is enough to remove it from the FS as shown in Fig\ref{q3}(b)-(c). The remaining pockets exhibit some anisotropy, which is enhanced with doping, and maximum DOS along $\Gamma$-M.

%\paragraph{Model - formalism.}

Having analyzed the input of our theory, we now discuss briefly (for details see the SOM) the technique that we employ to obtain our results. Our starting point is the fully anisotropic Eliashberg equations (e.g. see ref.\cite{Louie1}). The EPC that we use implies that the SC gap function is also separable and can be written as:
$\Delta_{{\bf k},n}=\Delta_n g_{\bf k}$, where $\Delta_n$ is the gap amplitude distributed over the Matsubara frequencies and $g_{\bf k}$ contains the momentum dependence of the SC gap ($|g_{\bf k}|\leq 1$). Utilizing this property, we end up with two eigenvalue equations at T=$T_c$ instead of one, as in standard Eliashberg theory. The first one is the usual eigenvalue problem but extended to account for the full anisotropy of the gap:
%\begin{eqnarray}\no
%\sum_{n'=0}^N \Bigl[a\cdot \left(\lambda(n-n')+\lambda(n+n'+1)\right)-2b -c\cdot\delta_{nn'}\\\label{tceq}
%\times\Bigl(\lambda(0)+2\sum_{m=1}^n\lambda(m)\Bigl)-\frac{\delta_{nn'}}{w_{n'}}\Bigl]\tilde{\Delta}_{n'}=0
%\end{eqnarray}
\begin{eqnarray}\no
\sum_{n'=0}^N \Bigl[a\cdot \left(\lambda(n-n')+\lambda(n+n'+1)\right)-2b -c\cdot\delta_{nn'}\\\label{tceq1}
\times\Bigl(\lambda(0)+2\sum_{m=1}^n\lambda(m)\Bigl)-\frac{\delta_{nn'}|\omega_{n'}|}{\pi T_c}\Bigl]\tilde{\Delta}_{n'}=0
\end{eqnarray}
where $\tilde{\Delta}_{n'}=\frac{\Delta_{n'}\pi T_c}{|\omega_{n'}|}$, $a=\frac{\langle g^*_{\bf k}\lambda_{{\bf kk}'}g_{{\bf k}'}\rangle_{{\bf kk}'}^{FS}}{\langle|g_{\bf k}|^2\rangle_{\bf k}^{FS}}$, $b=\frac{\mu^*|\langle g_{{\bf k}'}\rangle_{{\bf k}'}^{FS}|^2}{\langle|g_{\bf k}|^2\rangle_{\bf k}^{FS}}$, $c=\frac{\langle \lambda_{{\bf k}}|g_{\bf k}|^2\rangle_{{\bf k}}^{FS}}{\langle|g_{\bf k}|^2\rangle_{\bf k}^{FS}}$, $\lambda_{\bf k}=\langle\lambda_{{\bf k},{\bf k}'}\rangle_{{\bf k}'}^{FS}$ and even frequency pairing is explicitely assumed. The Coulomb pseudopotential, $\mu^*$, is taken as band independent.
The $T_c$ can be calculated by Eq.(\ref{tceq1}) provided that $g_{\bf k}$ is known. The latter can be obtained by the second eigenvalue equation:
\begin{eqnarray}\label{tceq2}
Z^{-1}_{\bf k}\langle\left(\lambda_{{\bf k},{\bf k}'}-\mu^*\right)g_{{\bf k}'}\rangle_{\bf k'}^{FS}=\rho g_{\bf k}
\end{eqnarray}
with $Z_{\bf k}=1+\lambda_{\bf k}$.
%and $\rho^{-1}$ $=\ln{\left[\frac{1.13 \omega_c}{T_c}\right]}$.
Eq.(\ref{tceq2}), when solved \textit{self-consistently}, yields the symmetry and the exact momentum dependence of the SC gap. This modelling, permits us to calculate accurately $g_{\bf k}$ with a large resolution of grid points and then obtain the exact $T_c$ within the full Eliashberg framework. The linearized equations do not provide information on the SC gap below $T_c$, e.g. whether the SC symmetry changes with T. In order to estimate such a possibility, we also calculate $g_{\bf k}$ at T=0 using the BCS-like self-consistent equation:
%by approximating Eq.\ref{El01} with $Z_{\bf k}=1+\lambda_{\bf k}$
%by neglecting $\Delta_{{\bf k},n}$ in Eq.\ref{El01} (e.g. see \cite{NicolCarbotte}), thus obtaining:
\begin{eqnarray}\label{BCSlike}
\Delta_{\bf k}=Z^{-1}_{\bf k}\langle \left(\lambda_{{\bf k},{\bf k}'}-\mu^*\right)\Delta_{{\bf k}'}\sinh^{-1}{\left(\omega_D/|\Delta_{{\bf k}'}|\right)}\rangle^{FS}_{\bf k'}
\end{eqnarray}
where $\omega_D$ is the Debye frequency.
%The corresponding square-well model free energy difference at T=0 is:
%\begin{eqnarray}\label{frwTz}
%\delta F=N_0\omega_c\langle Z_{\bf k}(\omega_c-\sqrt{\omega^2_c+|\Delta_{\bf k}|^2})\rangle^{FS}_{\bf k}
%\end{eqnarray}
The above, supplemented with the respective free energy formula, provides an approximate description for the gap structure at T=0 in the limit where the feedback of the gap on the renormalization function is negligible (for details see the SOM). Using the solution of Eq.(\ref{BCSlike}) in Eq.(\ref{tceq1}), we can calculate the anticipated $T_c$ of the T=0 state. A combination of the T=0 and T=$T_c$ results suffices in order to discern possible changes in the SC symmetry with T.
\begin{figure}[t!]
%\begin{flushright}
\vspace{0.18in}
\begin{minipage}[b]{1.68in}
\labellist
\footnotesize\hair 2pt
\scriptsize\pinlabel $T=T_c$ at 163 267
\footnotesize\pinlabel (a) at 25 238
\small\pinlabel $s_{++}$ at 100 200
\small\pinlabel $s_{\pm}$ at 180 110
%\small\pinlabel $p_{\pm}$ at 250 35
%\footnotesize\pinlabel $\mu$=0.00 at 75 255
%\footnotesize\pinlabel $\mu$=0.00 at 75 230
\footnotesize\pinlabel $\mu$=0.0 at 85 240
\small\pinlabel \rotatebox{90}{$q_c$\footnotesize $(a^{-1})$} at -10 -100
%\small\pinlabel \rotatebox{90}{$q_c$\footnotesize $(a^{-1})$} at 10 -100
\endlabellist
\includegraphics[width=\textwidth]{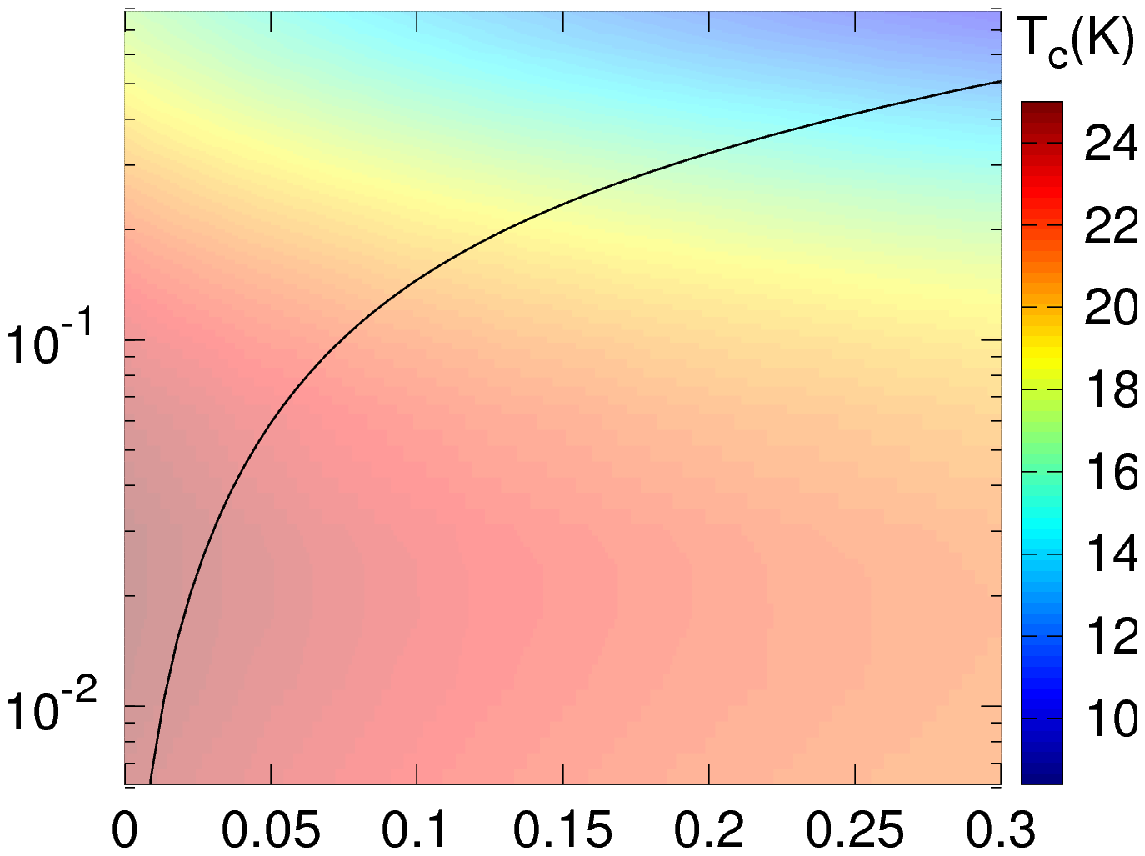}\vspace{0.02in}
%\vspace{0.15in}
\labellist
\footnotesize\hair 2pt
%\footnotesize\pinlabel (b) at 15 240
\footnotesize\pinlabel (b) at 25 238
\small\pinlabel $s_{++}$ at 120 190
\small\pinlabel $s_{\pm}$ at 210 168
\small\pinlabel $p_{\pm}$ at 190 80
%\footnotesize\pinlabel $\mu$=0.01 at 75 255
%\footnotesize\pinlabel $\mu$=0.01 at 75 230
\footnotesize\pinlabel $\mu$=10 at 80 235
\endlabellist
\includegraphics[width=\textwidth]{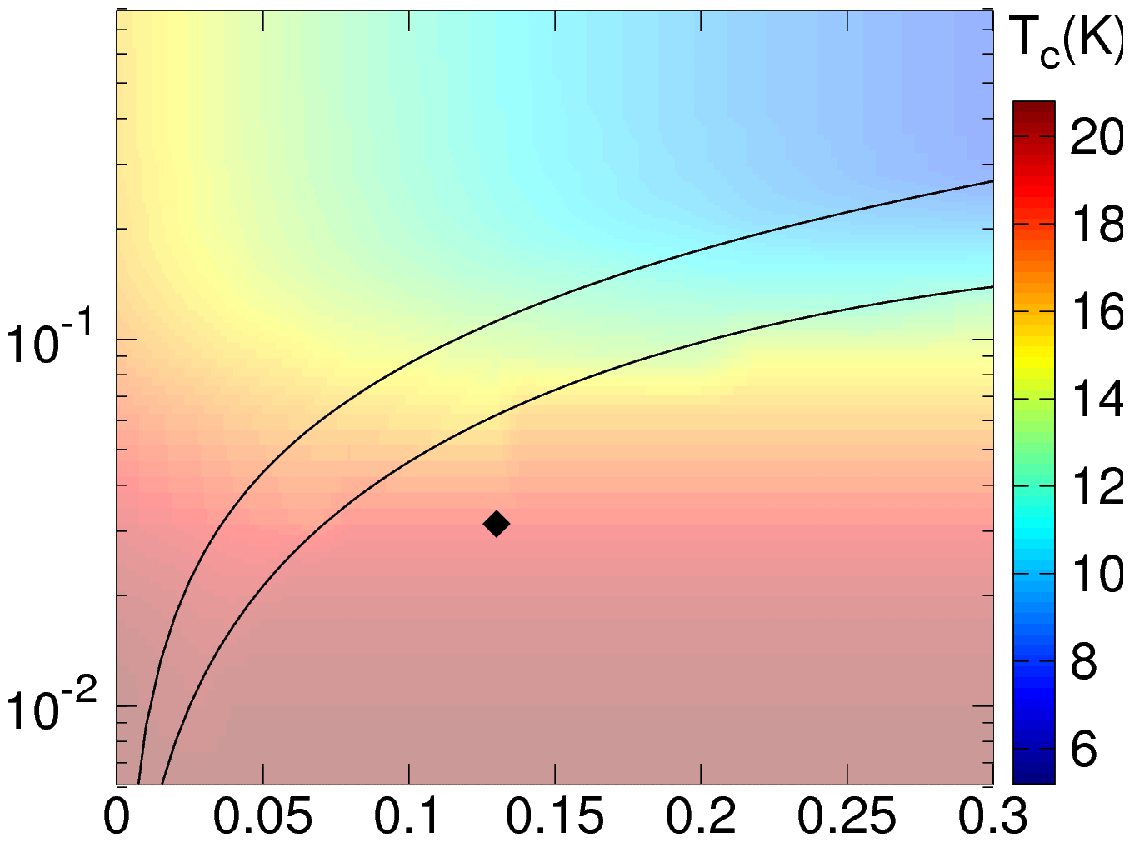}\vspace{0.02in}
\labellist
\footnotesize\hair 2pt
%\footnotesize\pinlabel (d) at 15 240
\footnotesize\pinlabel (c) at 25 238
\small\pinlabel $s_{++}$ at 105 190
\small\pinlabel $s_{\pm}$ at 185 165
\small\pinlabel $p_{\pm}$ at 200 100
%\footnotesize\pinlabel $\mu$=0.05 at 75 255
%\footnotesize\pinlabel $\mu$=0.05 at 75 230
\footnotesize\pinlabel $\mu$=50 at 80 238
\endlabellist
\includegraphics[width=\textwidth]{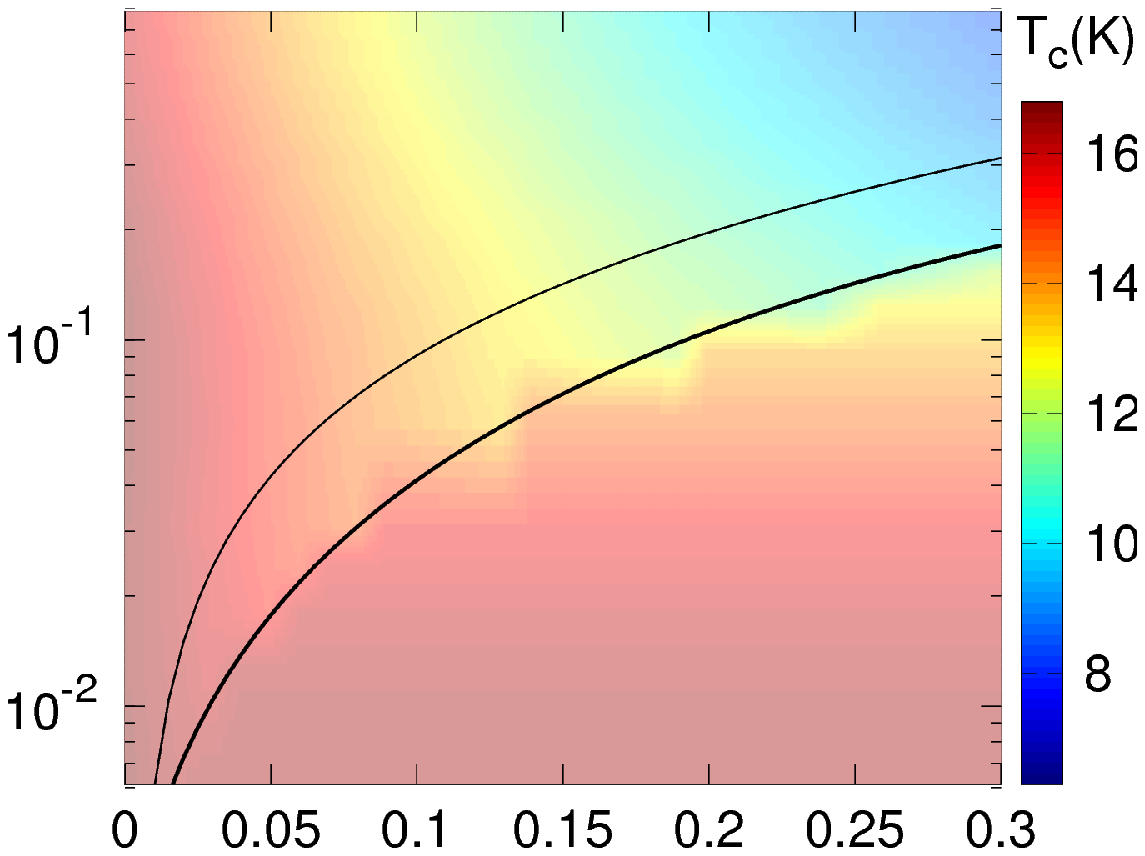}
%\vspace{0.05in}
\end{minipage}
\vspace{0.18in}
\begin{minipage}[b]{1.68in}
\labellist
\footnotesize\hair 2pt
\scriptsize\pinlabel $T=0$ at 163 267
\footnotesize\pinlabel (d) at 25 238
\small\pinlabel $s_{++}$ at 100 200
\small\pinlabel $s_{\pm}$ at 190 110
%\footnotesize\pinlabel $\mu$=0.00 at 75 255
%\footnotesize\pinlabel $\mu$=0.00 at 75 230
\footnotesize\pinlabel $\mu$=0.0 at 85 240
\endlabellist
\includegraphics[width=\textwidth]{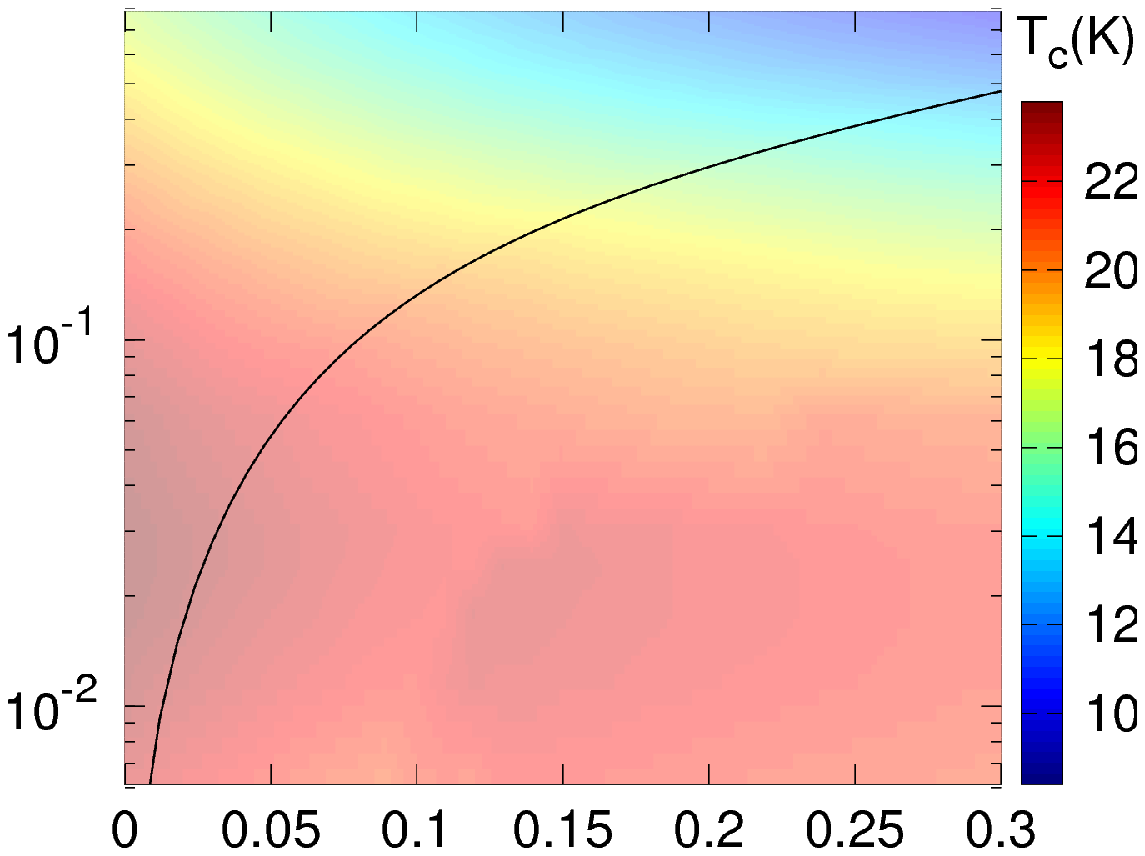}\vspace{0.02in}
%\vspace{0.15in}
\labellist
\footnotesize\hair 2pt
\footnotesize\pinlabel (e) at 25 238
\small\pinlabel $s_{++}$ at 120 190
\small\pinlabel $s_{\pm}$ at 200 120
\small\pinlabel $p_{\pm}$ at 230 50
%\footnotesize\pinlabel $\mu$=0.01 at 75 255
%\footnotesize\pinlabel $\mu$=0.01 at 75 230
\footnotesize\pinlabel $\mu$=10 at 80 235
\endlabellist
\includegraphics[width=\textwidth]{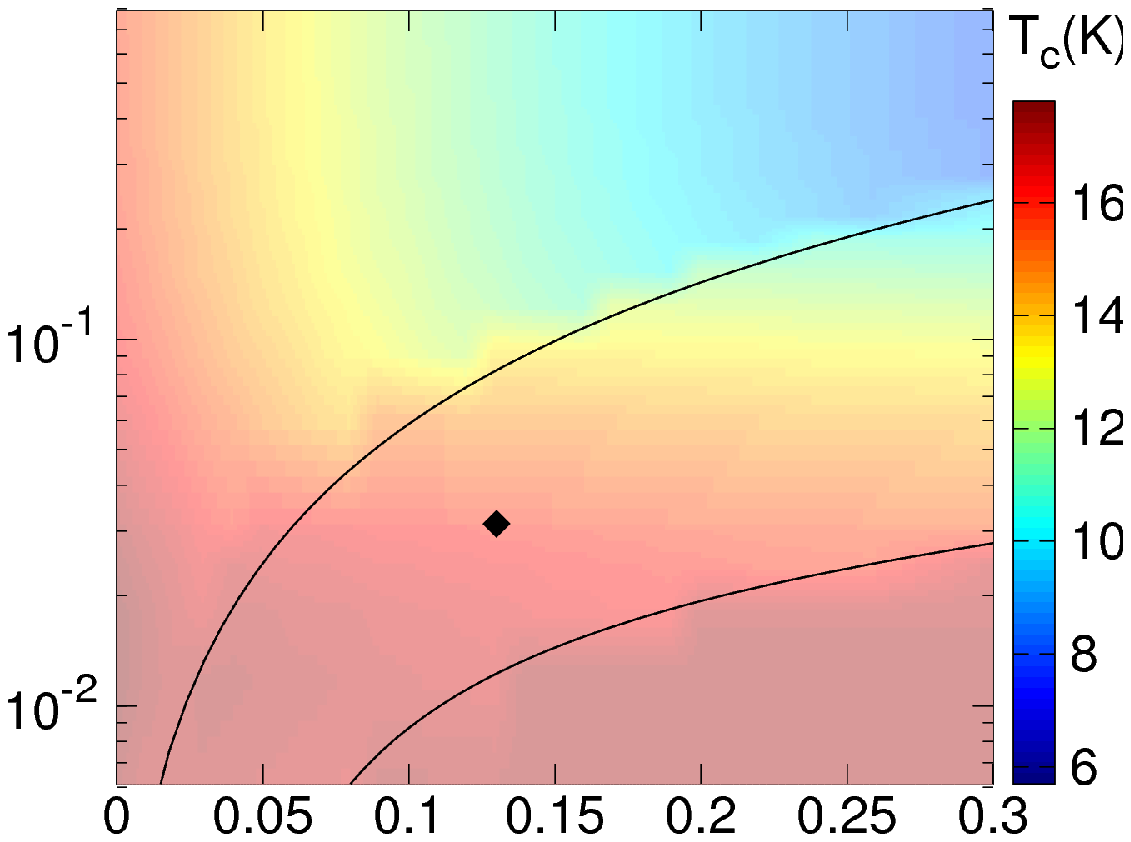}\vspace{0.02in}
\labellist
\footnotesize\hair 2pt
\footnotesize\pinlabel (f) at 25 238
\small\pinlabel $s_{++}$ at 115 190
\small\pinlabel $s_{\pm}$ at 215 165
\small\pinlabel $p_{\pm}$ at 220 100
%\footnotesize\pinlabel $\mu$=0.05 at 75 255
%\footnotesize\pinlabel $\mu$=0.05 at 75 230
\footnotesize\pinlabel $\mu$=50 at 80 238
%\footnotesize\pinlabel $\mu^*$ at -2 -20
\small\pinlabel $\mu^*$ at -2 -10
\endlabellist
\includegraphics[width=\textwidth]{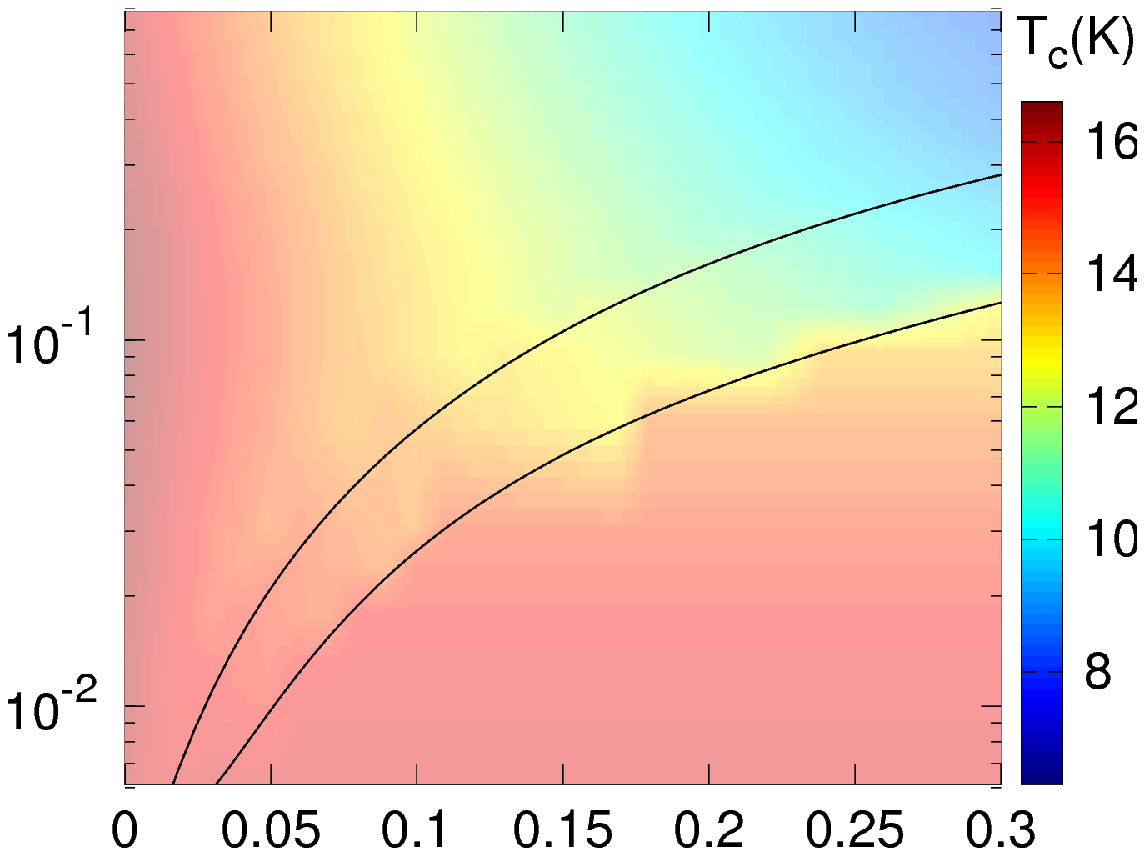}
%\vspace{0.05in}
\end{minipage}
%\end{flushright}
\caption{(Color online) (a)-(c) Self-consistently calculated phase diagrams as a function of the Coulomb pseudopotential $\mu^*$ and the EPI momentum cut-off $q_c$, for $\lambda$=$1.38$ and $\mu$=$[0,10,50]$meV at $T$=$T_c$. The y-axis is logarithmic to highlight the small-$q_c$ region. The background color grade corresponds to the calculated $T_c$ at each ($q_c$,$\mu^*$) value. (d)-(f) Same plots at $T$=$0$. Here, the color grade is for calculations of the \textit{anticipated} $T_c$ from the low-T momentum structure of the SC gap, as described in the text. The notation $s_{++},s_\pm$ and $p_\pm$ is for anisotropic one-sign and extended sign-alternating singlet $s$-wave and chiral triplet SC gap symmetries, respectively. It is easy to observe that the $\mu$=0 and $\mu$=50meV results for both T limits agree very well with each other. Hence. at these doping levels, we expect that the symmetry of the SC gap found for $T=T_c$ is unchanged down to zero T. On the contrary, for intermediate dopings (here $\mu$=$10meV$), there is a large parameter regime where the low-T gap symmetry is singlet $s_\pm$, while for higher temperatures, we find chiral triplet SC (e.g. compare the situation marked by the diamond symbols in Fig.(b) and (e)). These findings are in perfect agreement with the recently observed singlet to triplet SC temperature induced transitions in LiFeAs \cite{Baek2} (see Fig.4 for H=0).
%, where $p_\pm({\bf k})=\sin{k_x}\pm i\sin{k_y}$ is nodeless on the FS.
}\label{pdtc}
\end{figure}
We solve Eq.(\ref{tceq2}) \textit{self-consistently} over a 512$\times$512 grid of k-points for various initial forms of $g_{\bf k}$ including every allowed symmetry of the $D_{4h}$ group or random values. The solution that is kept, maximizes the eigenvalue $\rho$. Next, the coefficients $a$,$b$,$c$ are calculated and plugged into Eq.(\ref{tceq1}) which is solved for the $T_c$. The cutoff frequency used is $\omega_c=10\omega_{D}$, where for $\omega_D$ we use the calculated logarithmic frequency $\omega_{ln}$=100K \cite{Jishi}. These steps are repeated varying $\mu\in[0,50meV]$, $q_c\in[\frac{\pi}{512a},\frac{\pi}{4a}]$ (the lattice constant $a$=1) and $\mu^*\in [0,0.3]$ while keeping $\lambda$=1.38 fixed. The same procedure is followed for Eq.(\ref{BCSlike}) where the correct solution minimizes the free energy.

%\paragraph{Results-Discussion.}
The obtained $q_c$-$\mu^*$ phase diagrams (PD) and the calculated $T_c$ for various dopings at T=T$_c$ are shown in Fig.\ref{pdtc}(a)-(c). In the stoichiometric regime (Fig.\ref{pdtc}(a)), singlet A$_{1g}$ states dominate the PD. This behavior can be attributed to the respective FS topology of LiFeAs shown in Fig.\ref{q3}(a). Due to the VHp at $\Gamma$, electrons from the isotropic $\alpha$-band have a pronounced contribution to SC pairing thus favoring isotropic gap structures over SC symmetries with nodes such as d or p-wave. For relatively large $q_c$ values, we find a one-sign anisotropic, s$_{++}$, SC gap. Decreasing $q_c$ or increasing $\mu^*$ the system, in order to avoid the pair-breaking Coulomb repulsion, enters into a sign-alternating nodeless s$_\pm$ state. Even in the latter, the symmetry is not pure s$_\pm$, but a strong s-wave component is superimposed. Hence, the $T_c$ of both states is affected by $\mu^*$, since $b\neq 0$ in Eq.(\ref{tceq1}).
 Notice how $T_c$ increases with decreasing $q_c$ since then, the EPC becomes sharply peaked and tends to become purely intraband, maximizing the contribution to SC pairing. 
 
\begin{figure}[t!]
%\begin{center}
\begin{minipage}[b]{1.5in}
\labellist
\footnotesize\hair 2pt
\footnotesize\pinlabel a) at 15 210
\scriptsize\pinlabel $\mu$=0meV at 130 -12
\scriptsize\pinlabel $\Gamma$ at 8 8
\scriptsize\pinlabel X at 243 8
\scriptsize\pinlabel M at 243 215
\footnotesize\pinlabel $s_{++}$ at 125 125
\endlabellist
\includegraphics[width=\textwidth]{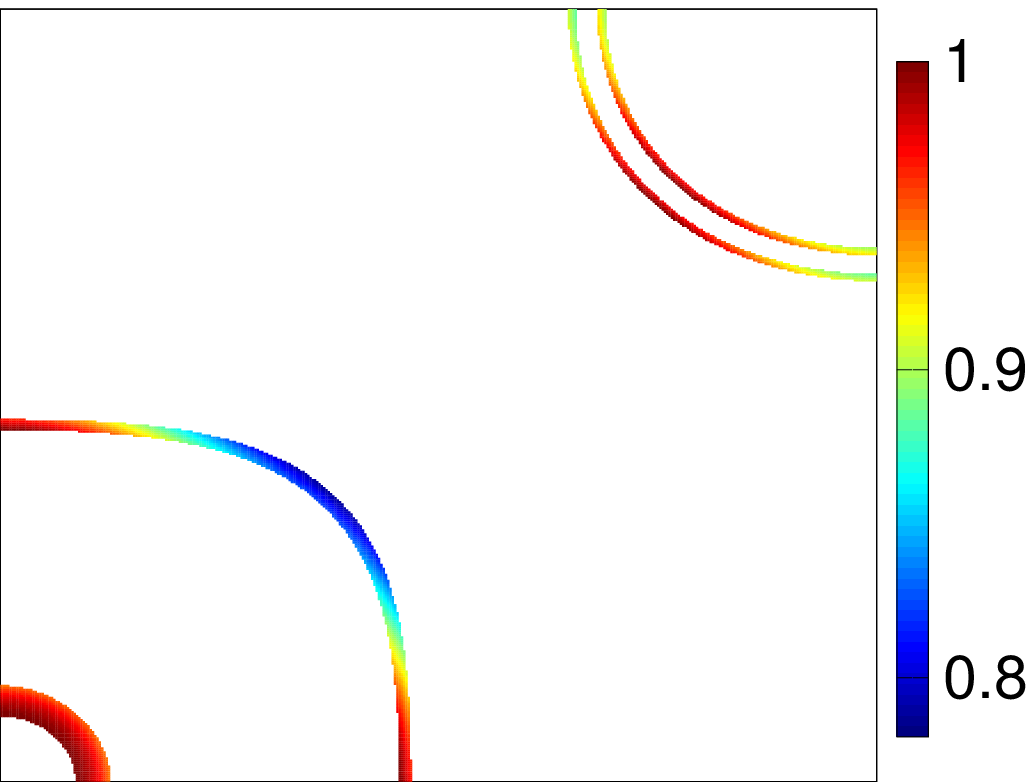}
\end{minipage}
\hspace{0.1in}
\begin{minipage}[b]{1.5in}
\labellist
\footnotesize\hair 2pt
\footnotesize\pinlabel b) at 15 210
\scriptsize\pinlabel $\mu$=10meV at 130 -12
\scriptsize\pinlabel $\Gamma$ at 8 8
\scriptsize\pinlabel X at 243 8
\scriptsize\pinlabel M at 243 215
\footnotesize\pinlabel $p_{\pm}$ at 125 125
\endlabellist
\includegraphics[width=\textwidth]{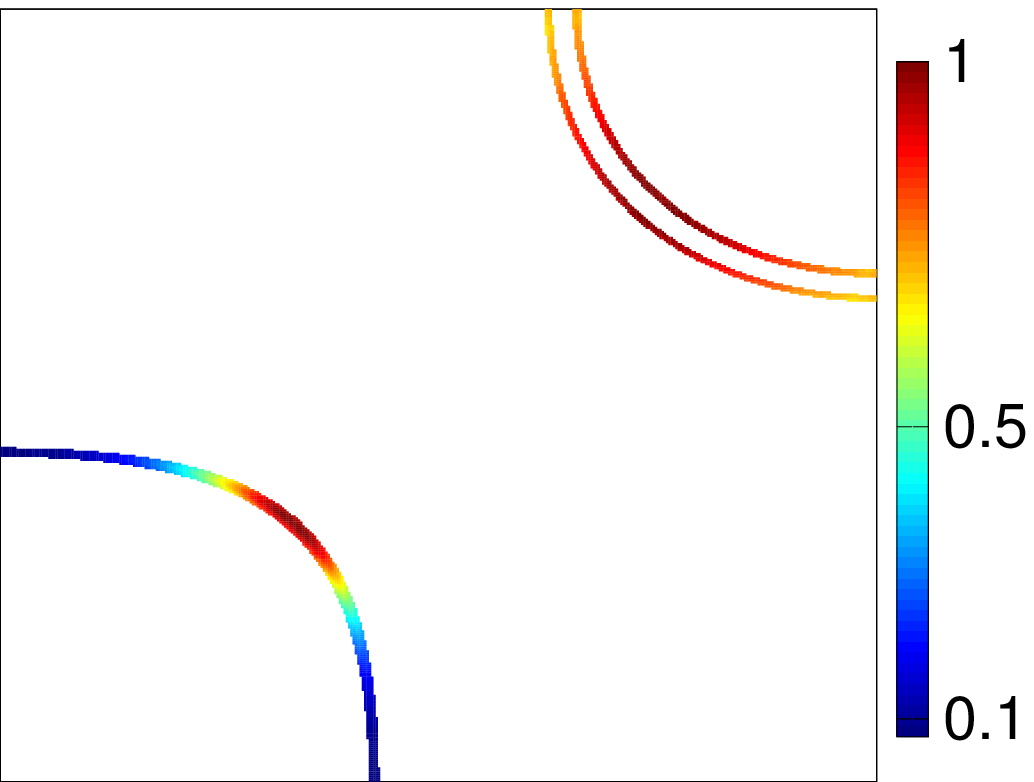}
\end{minipage}
%\end{center}
\caption{(Color online) Self-consistently obtained momentum dependence of the SC gap ($|g_{\bf k}|$) at $T$=$T_c$ for $\lambda$=1.38 and $\mu^*$=0.13 in the (a) stoichiometric $s_{++}$ regime ($\mu$=0meV, $q_c$=0.25) and (b) in the non-stoichiometric $p_\pm$ regime ($\mu$=10meV, $q_c$=0.032). The gap on each band is normalized to unity in order to highlight the location of gap minima/maxima. The $s_{++}$ SC gap in (a) has the exact FS momentum dependence reported by ARPES \cite{Bor1,Umezawa}.}\label{pdqcmu}
\end{figure}

 From the $T_c$ dependence, it can be seen that there is a reasonable parameter space where our phononic theory produces the experimental $T_c\approx$18K of LiFeAs. For this parameter space ($q_c>0.1$) we find that in both $s_{++}$ and $s_\pm$ regimes the gap over the $\alpha$-pocket is the largest and isotropic. This is a result of the combined VHp and MD effects. The gap on the other FS sheets shows anisotropy that varies depending on symmetry. In the $s_\pm$ state, the $\beta$-pocket exhibits maxima along the $\Gamma$-X and the electron pockets along the X-M direction, respectively. In the $s_{++}$ state, the $\beta$-pocket exhibits maxima along the $\Gamma$-X and the electron pockets along the $\Gamma$-M direction. The latter features match precicely the FS dependence reported by ARPES for this material \cite{Bor1,Umezawa}. For example, for $\mu^*=0.13$ and $q_c=0.25$ we find $s_{++}$ SC with $T_c\approx$18K and FS momentum dependence as shown in Fig.\ref{pdqcmu}(b).

In the non-stoichiometric case the $\alpha$-pocket is removed from the FS as can be seen in Fig.\ref{q3}(b)-(c) for $\mu=$10 and $\mu=$50 meV, respectively. The remaining anisotropic FS possesses minima along the directions parallel and perpendicular to $\Gamma$-X that are enhanced with $\mu$. Hence, the system tends to allow nodes or gap minima in these directions, possibly favoring $d_{xy}$ or p-wave SC. The respective T=$T_c$ PDs are shown in Fig.\ref{pdtc}(b)-(c). Increasing $\mu^*$ the $s_{\pm}$ region shrinks and at $q_c\lesssim 0.1$, \textit{a p-wave SC state} ($g_{\bf k}\sim\sin{k_x}$ or $\sin{k_y}$) is indeed stabilized. Since in our formalism SC pairing is intraband, this is necessary spin triplet in order for the antisymmetry of the electron wavefunction to be satisfied. Free energy calculations at T=0, indicate that the stable solution is in fact \textit{chiral} ($g_{\bf k}\sim \sin{k_x}\pm i\sin{k_y}$), which we hereafter refer to as $p_\pm$. The absence of the VHp affects also the $T_c$ which gradually decreases with $\mu$. For small $\mu$, this decrease can be compensated with a decrease in $q_c$, as is evident in Fig.\ref{pdtc}(b). Remarkably, for $(q_c,\mu^*)$ values that yield $T_c\approx$18K, the system is in the $p_\pm$ state. The occurence of $p_\pm$ SC \textit{only for $\mu>0$} within our theory is in perfect agreement with QPI \cite{QPItr} and particularly NMR \cite{Baek,Baek2} reports, where in the latter it has been explicitely associated with non-stoichiometry. Moreover, it resolves the conflict between the reported triplet SC and ARPES data that indicate a strong EPI in LiFeAs \cite{Kordyuk}.

We now compare our T=0 results with the ones at $T$=$T_c$. The T=0 PDs for $\mu$=$[0,10,50]$meV are shown in Fig.\ref{pdtc}(d)-(f). For $\mu$=0 (Fig.\ref{pdtc}(a) and (d)), it is easy to notice that not only the PDs but also the calculated $T_c$'s, in the parameter region of interest discussed above, agree nicely. The same also holds for the $\mu$=50meV results (Fig.\ref{pdtc}(c) and (f)). In fact, for $\mu>50$meV, the two PD's become exactly the same. Hence, we conclude that the T=0 approximation works very well both in determining the PD as well as predicting the $T_c$. On this firm basis we estimate that for $\mu$=0 ($\mu\geq$50meV) LiFeAs is in a $s_{++}$ ($p_\pm$) state down to zero T. 

Our solutions for $0<\mu<50$meV are amenable to an interesting interpretation. In this regime, we systematically find that for relevant ($q_c$,$\mu^*$) values, the SC symmetry at T=0 is $s_\pm$, while at T=$T_c$ it is $p_\pm$. Moreover, the expected $T_c$ for the $s_\pm$ is smaller than the $T_c$ of the $p_\pm$ state. Thus, we find clear evidence of \textit{a temperature induced transition from singlet $s_\pm$ to triplet $p_\pm$ SC that can take place for small deviations from stoichiometry in LiFeAs}. Remarkably, such transitions have been reported very recently by NMR in samples that are slightly non-stoichiometric \cite{Baek2}. For example, for $\mu$=10meV, $\mu^*$=0.13 and $q_c$=0.032, we find two successive transitions, to $p_\pm$ SC below $T_{c_1}\approx$18K and to $s_\pm$ SC below $T_{c_2}\approx$16K (see diamond symbol in Fig.\ref{pdtc}(b) and (e)), in perfect agreement with the findings of the latter experiments summarized in Fig.4 of ref.\cite{Baek2}.
% for H=0.
 Due to the small-$q_c$ values in this region, the enhanced MD leads to a SC gap structure whose FS momentum dependence follows closely the FS DOS. Thus, we predict that the SC gap in both the $s_\pm$ and $p_\pm$ regions should exhibit maxima along $\Gamma$-M, as is shown e.g. in Fig.\ref{pdqcmu}(c). Our key results for $\mu^*$=0.13 are summarized on Table \ref{table1}.

Finally, given the close proximity of $s_{++}$, $s_\pm$ and $p_\pm$ states found within our theory, it is natural to expect that an applied magnetic field would induce a singlet to triplet transition in LiFeAs, either by suppresing singlet pairing or by enhancing the FS DOS anisotropy or both, thus favoring $p_\pm$ SC. Hence, our mechanism could provide 
a plausible explanation for the experimental reports of a field 
induced triplet SC phase in this material \cite{Baek,lifeasField,Baek2}.
{\small \setlength{\tabcolsep}{8pt}
\begin{table}[t]
\caption{Summary of key results for $\lambda$=1.38, $\mu^*$=0.13}
\begin{tabular}{cccccc}
%\hline
\hline\\ [-2.3ex]
$\mu$ (meV)&$q_c$&$T_{c_1}$ (K)&$g_1({\bf k})$&$T_{c_2}$ (K)&$g_2({\bf k})$\\ [0.5ex]
\hline
$0$&0.25&$18.2$&$s_{++}$&-&-\\
$10$&0.042&18&$p_{\pm}$&15.6&$s_\pm$\\
$50$&0.032&15.5&$p_{\pm}$&-&-\\\hline
\end{tabular}\label{table1}
\end{table}}

%\paragraph{Conclusions}

In conclusion, we have presented a theory from the small-q phonon perspective, within which a plethora of controversial and yet unexplained experimental findings in SC LiFeAs are coherently understood. Based on realistic, fully anisotropic Eliashberg calculations, we have demonstrated how the interplay of this small-q EPI and the intrinsic FS properties of LiFeAs, governed by the presence of a VHp at the center of the BZ, 
result in a delicate balance between singlet $s_{++}$,$s_\pm$ and triplet chiral $p_x+ip_y$ SC. Furthermore, we explicitly showed that slight deviations from stoichiometry and/or changes in temperature may favor one of these states. Our results resolve the conflict between a strong EPC and one sign singlet SC reported by ARPES and unconventional and, most interestingly, chiral triplet SC observed by other probes. In addition, they provide a systematic understanding for the occurence of triplet SC in non stoichiometric samples as well as the temperature induced singlet to triplet SC transitions reported very recently.
The accuracy of the present results and the exotic character of the 
involved SC states,
establish that the small-q EPI 
mechanism should be considered on an \emph{equal footing} with
 spin and other purely electronic mechanisms,
in the analysis of \emph{any} unconventional superconductor.
Clearly, even the most exotic SC states 
like the chiral spin triplet reported here, should not exclude apriori 
%the involvement of phonons.
our phononic mechanism.

\begin{acknowledgments}
We are grateful to S.Borisenko and B.B\"uchner for enlightening discussions. We also thank S.Borisenko for providing us the ARPES TB fit for LiFeAs prior to its publication. A.A. acknowledges financial support by $\Pi$EBE of National Technical University of Athens.
\end{acknowledgments}

\begin{widetext}
\vspace{8 mm}
{\bf\Large \center Supplementary Online Material for ``Phonon mediated singlet and chiral spin triplet superconductivity in LiFeAs''} \\
\setcounter{equation}{0}
\renewcommand{\theequation}{S\arabic{equation}}
\setcounter{figure}{0}
\renewcommand{\thefigure}{S\arabic{figure}}

%\title{SUPPLEMENTARY ONLINE MATERIAL\\for\\
%Phonon mediated singlet and chiral spin triplet superconductivity in LiFeAs}
%\author{A. Aperis}\email{aaperis@mail.ntua.gr}
%\author{G. Varelogiannis}
%\affiliation{Department of Physics, National Technical University of Athens, GR-15780 Athens, Greece}
%\vskip 0.7cm
%\maketitle
%\setcounter{equation}{0}

\section{Small-q electron-phonon coupling and its separability}

In this section, we discuss the simplest situation when a small-q electron-phonon coupling (EPC) may arise and show how the obtained EPC maps to a sebarable one without affecting any of the basic physics. The EPC function is defined as:
\begin{eqnarray}\label{lkknn}
\lambda({\bf k},{\bf k}';n,n')=2\int_0^\infty \frac{\alpha^2F({\bf k},{\bf k}';\Omega)\Omega}{(\omega_n-\omega_{n'})^2+\Omega^2}d\Omega
\end{eqnarray}
where $\alpha^2F({\bf k},{\bf k}';\Omega)$ is the momentum dependent electron-phonon spectral function (EPSF), which for a single phonon branch is:
\begin{eqnarray}\label{EPSFkkw}
\alpha^2F({\bf k},{\bf k}';\Omega)=N_0|g_{{\bf kk}'}|^2\delta\left(\Omega-\omega_{{\bf k}-{\bf k}'}\right)
\end{eqnarray}
with $g_{{\bf kk}'}$
%=\sqrt{\frac{\hbar}{2M\omega_{{\bf q},j}}}\bra{{\bf k}}\hat{\epsilon}_{{\bf q},j}\cdot\nabla V\ket{{\bf k+q}}$
the electron-phonon matrix element (EPME) \cite{Grimvall}.
%\subsection{A simple model of small-q EPC}
Perhaps the most common situation when small-q processes dominate can take place when large dielectricity and the long range nature of the Coulomb potential are considered, as first discussed by Abrikosov (ref. \cite{smq1} of main text). As a paradigm, we analyze this case in some detail. Due to the modulation of the screened Coulomb potential, there exists a first-order coupling between electrons and phonons \cite{Schrieffer}. In this case, the electron-phonon matrix element is given by:
\begin{equation}
g_{\bf q}=\sqrt{\frac{\hbar}{2M\omega_{{\bf q}}}}\hat{\epsilon}_{{\bf q}}\cdot{\bf q}\frac{V_{\bf q}}{\epsilon_{\bf q}}
\end{equation}
where ${\bf q}={\bf k}-{\bf k}'$ is the exchanged momentum, M is the ionic mass, $\omega_{\bf q}$ is the renormalized phonon dispersion, $\hat{\epsilon}_{{\bf q}}$ is the phonon polarization, $V_{\bf q}=\frac{4\pi e^2}{\epsilon q^2}$ is the bare Coulomb potential, $\epsilon$ is the static dielectric constant and $\epsilon_{\bf q}$ is the static dielectric function. The renormalized phonon dispersion is just $\omega_{\bf q}=\frac{\Omega_p}{\sqrt{\epsilon_{\bf q}}}$, where $\Omega_p$ is the phonon plasma frequency. In the simplest picture, the dielectric function can be written in the Thomas-Fermi approximation as: $\epsilon_{\bf q}=1+\frac{q^2_{TF}}{q^2}$, where $q_{TF}$ is the Thomas-Fermi screening wavevector and $q_{TF}=\sqrt{4\pi e^2 N_0/\epsilon}$.

Compining the above relations, the EPC can be written in the following form:
\begin{eqnarray}\label{lkk1}
\lambda({\bf q};m)=N_0 V_{ep}\frac{1}{q^2+q^2_{TF}}\frac{\frac{q^2\Omega^2_p}{q^2+q^2_{TF}}}{w_m^2+\frac{q^2\Omega^2_p}{q^2+q^2_{TF}}}
\end{eqnarray}
where the effective parameter $V_{ep}$ includes all remaining terms and $w_m=2m\pi T$ is a bosonic frequency ($m=n-n'$). The EPC has a small-q momentum dependence peaked at $q=q_{TF}$ and a Lorentzian frequency dependence whose width is determined again by $q_{TF}$. Thus, even in this rough approximation, the EPC acquires a small-q momentum dependence with the momentum cutoff defined by the Thomas-Fermi screening wavevector. The peak at small-q gets more pronounced by reduction of the density of states or an increase in the dielectric constant.
At low energies, ($w$=0 or T=0), Eq.(\ref{lkk1}) reduces to:
\begin{eqnarray}\label{lkk0}
\lambda({\bf q})=N_0V_{ep}\frac{1}{q^2+q^2_{TF}}
\end{eqnarray}
Increasing the ratio T/$\Omega_p$, Eq.(\ref{lkk1}) approaches the form of Eq.(\ref{lkk0}) for $m\neq 0$. For parameters relevant to LiFeAs this is the case near $T\approx T_c$ (i.e. see Fig.\ref{smqsep}). Thus, $\lambda_{{\bf q},m}$ behaves like a separable function over momentum and frequency in the physical range that we are interested in.

\subsection{Separable small-q EPC}

Let us now assume an effective separable EPC. This can be achieved, for example, if the EPSF is separable and the frequency dependent part is described by an Einstein spectrum \cite{DaamsCarbotte,MillisIn}:
\begin{eqnarray}\label{Eph}
\alpha^2_EF({\bf k,k'};\Omega)=\lambda_{\bf kk'}\frac{\Omega_E}{2}\delta(\Omega-\Omega_E)
\end{eqnarray}
Inserting Eq.(\ref{Eph}) into Eq.(\ref{lkknn}) one easily gets:
\begin{eqnarray}\label{lkknnEph}
\lambda({\bf k},{\bf k}';n,n')=\lambda_{{\bf kk}'}\frac{\Omega_E^2}{(\omega_n-\omega_{n'})^2+\Omega_E^2}=\lambda_{{\bf kk}'}\lambda_{nn'}
\end{eqnarray}
At low energies ($n=n'$/T=0) the above reduces to: $\lambda_{\bf kk'}$ which we assume to have a small-q structure within a cutoff $q_c$. Thus, we can write:
\begin{eqnarray}\label{lkknnsep}
\lambda_{sep}({\bf q};m)=N_0 V_{ep}\frac{1}{q^2+q^2_c}\frac{\Omega_E^2}{w_m^2+\Omega_E^2}
\end{eqnarray}
Obviously, the above matches exactly Eq.(\ref{lkk1}) at low energies ($w$=0/T=0) and for $q_c=q_{TF}$, $\Omega_p=\Omega_E$. In our calculations for LiFeAs, we set $\Omega_{p(E)}$=$\omega_{log}$=100 K (ref. \cite{Jishi} of main text). We observe that already at T=10K the EPC of Eq.(\ref{lkknn}) and Eq.(\ref{lkknnsep}) agree very well as it is shown in Fig.\ref{smqsep}.
\begin{figure}[h!]
\includegraphics[width=0.4\textwidth]{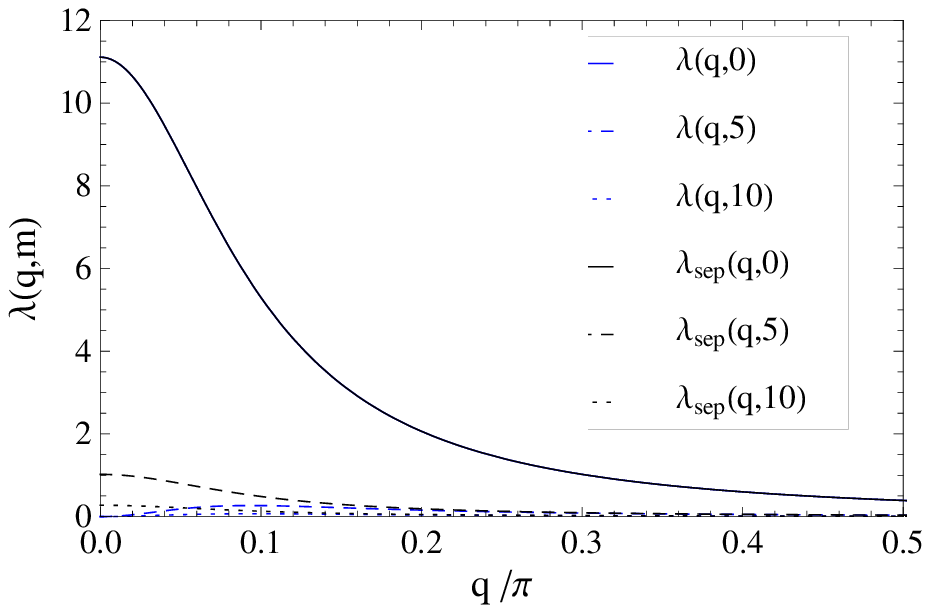}
\caption{Momentum dependence of $\lambda({\bf q},m),\lambda_{sep}({\bf q},m)$ for $q_{TF(c)}$=0.3, $\Omega_{p(E)}$=100K at T=10K and different Matsubara frequencies. The agreement between $\lambda$ and $\lambda_{sep}$ is evident.}\label{smqsep}
\end{figure}

\section{Multiband momentum dependent Eliashberg formalism}

Our starting point is the system of coupled anisotropic Eliashberg equations at the finite temperature:
\begin{eqnarray}\label{El01}
Z_{{\bf k},n}&=&1+\frac{\pi T}{\omega_n}\sum_{n'}\langle\lambda({\bf k},{\bf k}';n,n')\frac{\omega_{n'}}{\sqrt{\omega^2_{n'}+\Delta_{{{\bf k}',n'}}^2}}\rangle^{FS}_{{\bf k}'}\\\label{El02}
Z_{{\bf k},n}\Delta_{{\bf k},n}&=&\pi T\sum_{n'}^{|\omega_{n'}|<\omega_c}\langle\Bigl[\lambda({\bf k},{\bf k}';n,n')-\mu^*(\omega_c)\Bigl]
\frac{\Delta_{{\bf k}',n'}}{\sqrt{\omega^2_{n'}+\Delta_{{{\bf k}',n'}}^2}}\rangle^{FS}_{{\bf k}'}
\end{eqnarray}
where the nth Matsubara freuency $\omega_n=\left(2n+1\right)\pi T$, $Z_{{\bf k},n}=Z({\bf k},i\omega_n)$ is the strong-coupling renormalization parameter, $\Delta_{{{\bf k},n}}=\Delta({{\bf k},i\omega_{n}})$ is the SC gap function, $\lambda({\bf k},{\bf k}';n,n')$ the EPC and $\mu^*(\omega_c)$ is the Coulomb pseudopotential which comes with a cutoff $\omega_c$ in the Matsubara frequency summation and is taken as band independent. The notation $\langle\ldots\rangle_{\bf k}^{FS}=\sum_{\bf k}\frac{N_{\bf k}}{N_0}\left(\dots\right)$ means an average over the entire FS, where $N_0$ is the Density of States (DOS), $N_{\bf k}=|\nabla \xi_{{\bf k}_F}|^{-1}$ the angularly resolved DOS (arDOS) and $\xi_{\bf k}$ the energy dispersion at the Fermi level ($\sum_{\bf k}^{{\bf k}\in FS}\frac{N_{\bf k}}{N_0}=1$). In this formalism, all band structure effects are conveniently encoded in the FS averages and the anisotropy of the FS is fully taken into account. 

Eq.(\ref{El01})-(\ref{El02}) provide the strong coupling description of a phonon mediated, singlet or unitary triplet, superconductor (e.g. see \cite{Allen}) having an arbitrary number of bands contributing to the FS \cite{Louie2}. In this sense, $\Delta_{{\bf k},n},Z_{{\bf k},n}$ are global quantities defined over the entire FS. For an $i$-band SC, it is possible to degrade the above into a system of $2i$-coupled equations for $\Delta^i_{{\bf k},n},Z^i_{{\bf k},n}$ defined on each separate band, however by doing so, the momentum dependence of the EPC is lost (e.g. see ref.\cite{Louie2}). Thus, throughout this study, we work in the most generalized framework as defined above. Note that, since the values of $\mu$ that we consider are very small, we have neglected the energy shift self-energy term.

\subsection{Equations at $T_c$ in the separable model}

%The above system can be linearized at $T$=$T_c$ to give:
%\begin{eqnarray}\label{anEliashZ}
%Z_{{\bf k},n}&=&1+w_n\sum_{n'}\langle\lambda({\bf k},{\bf k}';n,n')\rangle_{\bf k'}^{FS}s_n s_{n'}\\\no
%Z_{{\bf k},n}\Delta_{{\bf k},n}&=&\sum_{n'}^{|\omega_{n'}|<\omega_c}w_{n'}\langle\left\{\lambda({\bf k},{\bf k}';n,n')-\mu^*(\omega_c)\right\}\\\label{anEliashD}
%&\times&\Delta_{{\bf k}',n'}\rangle_{\bf k'}^{FS}
%\end{eqnarray}
%where $w_n=\frac{1}{|2n+1|}$ and $s_n=\text{sgn}(\omega_n)=\text{sgn}(n+\frac{1}{2})$. Introducing the separable approximation: $\lambda({\bf k},{\bf k}';n,n')=\lambda_{{\bf k},{\bf k}'}\lambda_{n,n'}$, Eq.(\ref{anEliashZ}) yields:
Linearizing the above system at $T_c$, introducing the separable approximation: $\lambda({\bf k},{\bf k}';n,n')=\lambda_{{\bf k},{\bf k}'}\lambda_{n,n'}$ and re-expressing the Matsubara sum on positive frequencies yields:
\begin{eqnarray}\label{Zlin}
Z_{{\bf k},n}&=&1+\lambda_{\bf k}\frac{1}{2n+1}\Bigl(\lambda(0)+2\sum_{m=1}^n\lambda(m)\Bigl)\\\label{Dlin}
Z_{{\bf k},n}\Delta_{{\bf k},n}&=&\sum_{n'=0}^{\omega_{n'}<\omega_c}w_{n'}\langle\Bigl\{\lambda_{{\bf k},{\bf k}'}\left(\lambda_{n-n'}+\lambda_{n+n'+1}\right)-2\mu^*(\omega_c)\Bigl\}\Delta_{{\bf k}',n'}\rangle_{\bf k'}^{FS}
\end{eqnarray}
where we have explicitely assumed even frequency SC and $\lambda_{\bf k}$=$\langle\lambda_{{\bf k},{\bf k'}}\rangle^{FS}_{\bf k'}$, $w_n$=$\frac{1}{2n+1}$,  and we have used the relation: $\sum_{n'}\lambda_{nn'}\frac{\omega_{n'}}{|\omega_{n'}|}=\lambda(0)+2\sum_{m=1}^n\lambda(m)$. From Eq.(\ref{Zlin}), it is evident that $Z_{{\bf k},n}$ has become separable. Inserting Eq.(\ref{Zlin}) into Eq.(\ref{Dlin}) we see that the SC gap has also acquired a separable structure and thus, we can write: $\Delta_{{\bf k},n}$=$\Delta_n g_{\bf k}$, where $\Delta_n$ contains the frequency and $|g_{\bf k}|\leq 1$ the momentum dependence, respectively. 

Inserting Eq.(\ref{Zlin}) into Eq.(\ref{Dlin}), multiplying both sides with $g^i_{\bf k}=\frac{g^*_{\bf k}}{\langle|g_{\bf k}|^2\rangle_{\bf k}^{FS}}$ and then taking the FS average gives:
%\widetext
%\begin{eqnarray}
%\Delta_n g_{\bf k}=\sum_{n'=0}^{\omega_{n'}<\omega_c}\left[w_{n'}\langle\Bigl\{\lambda_{{\bf k},{\bf k}'}\left(\lambda_{n-n'}+\lambda_{n+n'+1}\right)-2\mu^*(\omega_c)\Bigl\}g_{{\bf k}'}\rangle_{\bf k'}^{FS}\Delta_{n'}-g_{\bf k}\lambda_{\bf k}\delta_{nn'} w_{n'}\Bigl(\lambda(0)+2\sum_{m=1}^{n'}\lambda(m)\Bigl)\Delta_{n'}\right]
%\end{eqnarray}
\begin{eqnarray}\no
\Delta_n =\sum_{n'=0}^{\omega_{n'}<\omega_c}\Biggl[\Bigl\{\frac{\langle g^*_{\bf k}\lambda_{{\bf k},{\bf k}'}g_{{\bf k}'}\rangle_{\bf k,k'}^{FS}}{\langle|g_{\bf k}|^2\rangle_{\bf k}^{FS}}\left(\lambda_{n-n'}+\lambda_{n+n'+1}\right)
-2\mu^*(\omega_c) \frac{|\langle g_{\bf k'}\rangle_{\bf k'}^{FS}|^2}{\langle|g_{\bf k}|^2\rangle_{\bf k}^{FS}}\Bigl\}
-\frac{\langle|g_{\bf k}|^2\lambda_{\bf k}\rangle_{\bf k}^{FS}}{\langle|g_{\bf k}|^2\rangle_{\bf k}^{FS}}\delta_{nn'}\\ \times\Bigl(\lambda(0)+2\sum_{m=1}^{n'}\lambda(m)\Bigl)\Biggl]w_{n'}\Delta_{n'}
\end{eqnarray}
which, after some rearanging can be written as:
\begin{eqnarray}\no
\sum_{n'=0}^N \Bigl[a\cdot \left(\lambda(n-n')+\lambda(n+n'+1)\right)-2b -c\cdot\delta_{nn'}
\times\Bigl(\lambda(0)+2\sum_{m=1}^n\lambda(m)\Bigl)-\frac{\delta_{nn'}}{w_{n'}}\Bigl]\tilde{\Delta}_{n'}=0
\end{eqnarray}
with $\tilde{\Delta}_{n'}=\Delta_{n'}w_{n'}$, $a=\frac{\langle g^*_{\bf k}\lambda_{{\bf kk}'}g_{{\bf k}'}\rangle_{{\bf kk}'}^{FS}}{\langle|g_{\bf k}|^2\rangle_{\bf k}^{FS}}$, $b=\frac{\mu^*|\langle g_{{\bf k}'}\rangle_{{\bf k}'}^{FS}|^2}{\langle|g_{\bf k}|^2\rangle_{\bf k}^{FS}}$ and $c=\frac{\langle \lambda_{{\bf k}}|g_{\bf k}|^2\rangle_{{\bf k}}^{FS}}{\langle|g_{\bf k}|^2\rangle_{\bf k}^{FS}}$. The above is an eigenvalue problem similar to the one encountered in usual Eliashberg theory \cite{AllenDynes} but extended to include the full momentum dependence of the SC gap and it can be solved following standard methods \cite{Allen} for the $T_c$ provided that $g_{\bf k}$ is known. Obviously, for $g_{\bf k}=1$ the usual isotropic Eliashberg equation is retrieved.

Since $\Delta_{{\bf k},n}=\Delta_n g_{\bf k}$, we can apply a square-well ansatz \cite{CarbotteRev} to Eq.(\ref{Zlin})-(\ref{Dlin}) in order to extract an equation for the momentum dependent part of the SC gap:
\begin{eqnarray}\no
\lambda({\bf k},{\bf k}';n,n')=\begin{cases}\lambda_{{\bf k},{\bf k}'}\ \text{for both}\ |\omega_n |,|\omega_{n'}|<\omega_c \  ,\\
0 \ \text{otherwise}\end{cases}
\text{and}\ \ \ \ \ 
\Delta_n g_{\bf k}=\begin{cases}\Delta(T)g_{\bf k}\ ,\ |\omega_n |<\omega_c \  ,\\
0\ , \ |\omega_n|>\omega_c\end{cases}
\end{eqnarray}
It is easy to observe that within this model $Z_{{\bf k},n}$ becomes:
\begin{eqnarray}\label{anBCSZ}
Z_{\bf k}=1+\lambda_{\bf k}\end{eqnarray}
and the gap equation takes the form:
\begin{eqnarray}\no
&&Z_{\bf k}g_{\bf k}\Delta(T)=\sum_{n'=0}^{|\omega_{n'}|<\omega_c}w_{n'}\langle\left\{2\lambda_{{\bf k},{\bf k}'}-2\mu^*(\omega_c)\right\}g_{{\bf k}'}\rangle_{\bf k'}^{FS}\Delta(T)
\end{eqnarray}
which after a standard summation on Matsubara frequencies yields:
\begin{eqnarray}\label{tcgk}
Z^{-1}_{\bf k}\langle\left(\lambda_{{\bf k},{\bf k}'}-\mu^*\right)g_{{\bf k}'}\rangle_{\bf k'}^{FS}=\rho g_{\bf k}
\end{eqnarray}
where $\rho^{-1}=\ln{\left[\frac{1.13 \omega_c}{T_c}\right]}$. This is an eigenvalue equation for the momentum part of the SC gap alone, which when solved self-consistently provides the exact $g_{\bf k}$. Notice that in this case $g_{\bf k}$ is not just a form factor corresponding to an irreducible representation, but it is rather a superposition of form factors and all their harmonics of all the allowed irreducible representations, i.e. it is the \emph{realistic} SC structure favored by the system's specific characteristics.

\subsection{Equations at $T$=0 in the separable model}

At T=0, the Eliashberg equations can be treated approximately to give a closed form, BCS-like equation. This can be achieved by neglecting $\Delta_{{\bf k},n}$ in Eq.(\ref{El01}) and further applying the square-well ansatz (see e.g. ref.\cite{NicolCarbotte,Dolgov}). Using $Z_{\bf k}=1+\lambda_{\bf k}$ in Eq.(\ref{El02}) and  taking the zero T limit ($T\sum_{n'}\rightarrow2\int_0^{\omega_c}\frac{d\omega}{2\pi}$) gives:

\begin{eqnarray}\label{BCSlike}
\Delta_{\bf k}=Z^{-1}_{\bf k}\langle \left(\lambda_{{\bf k},{\bf k}'}-\mu^*\right)\Delta_{{\bf k}'}\sinh^{-1}{\left(\frac{\omega_c}{|\Delta_{{\bf k}'}|}\right)}\rangle^{FS}_{\bf k'}
\end{eqnarray}

After self-consistently solving the above equation, $g_{\bf k}$ is retrieved. The correct solution is determined by minimizing the respective free energy difference between the normal and the SC state within Eliashberg theory \cite{dfrBard}. A momentum dependent expression for the latter is \cite{dfr}:
\begin{eqnarray}\no
\delta F =-\pi T N_0\sum_n\langle  \left(\sqrt{\omega^2_n+\Delta^2_{{\bf k},n}}-|\omega_n|\right)
\Bigl(Z_{{\bf k},n}-Z^N_{{\bf k},n}\frac{|\omega_n|}{\sqrt{\omega^2_n+\Delta^2_{{\bf k},n}}}\Bigl) \rangle_{\bf k}^{FS}
\end{eqnarray}
Applying the square-well approximation, taking the T=0 integral over Matsubara frequencies and for $Z^N_{\bf k}=Z_{\bf k}$ we find the respective condensation energy as:
\begin{eqnarray}\label{dfrT0}
\delta F=N_0\omega_c\langle  Z_{\bf k}\bigl(\omega_c-\sqrt{\omega_c^2+\Delta^2_{{\bf k}}}\bigl)\rangle_{\bf k}^{FS}
\end{eqnarray}
%Eq.(\ref{BCSlike})-(\ref{dfrT0})
The isotropic version of Eq.(\ref{BCSlike}) has been previously considered in multiband Eliashberg theories, where it has been shown that it, in fact, captures the full Eliashberg results very well (e.g. see ref.\cite{Dolgov}). Our results for LiFeAs, indicate that this approximation scheme works also well, when the full momentum dependence is also included in the calculations.

%\begin{figure}[h!]
%%\begin{center}
%\begin{minipage}[b]{3.4in}
%\labellist
%\footnotesize\hair 2pt
%%\footnotesize\pinlabel (d) at 65 420
%%\footnotesize\pinlabel $h_1$ at 170 220
%\tiny\pinlabel $\Gamma$ at 130 80
%\tiny\pinlabel X at 240 80
%\tiny\pinlabel $\Gamma$ at 350 80
%\tiny\pinlabel M at 460 80
%\tiny\pinlabel $\Gamma$ at 570 80
%\tiny\pinlabel \rotatebox{90}{$E({\bf k})$ (meV)} at 70 280
%\footnotesize\pinlabel $\alpha$ at 160 220
%\footnotesize\pinlabel $\beta$ at 170 390
%\footnotesize\pinlabel $\gamma$ at 430 390
%\footnotesize\pinlabel $\delta$ at 520 390
%\endlabellist
%%\includegraphics[width=0.4\textwidth, height=0.2\textwidth]{hsplot}
%\includegraphics[width=0.65\textwidth, height=0.3\textwidth]{hsplot}
%\end{minipage}
%%\end{center}
%\caption{The energy dispersion of the four bands along high symmetry lines for $\mu=0$ \cite{}.}\label{q3}
%\end{figure}

\end{widetext}

\end{document}